\documentstyle[aps,12pt]{revtex}
\begin{document} 
\draft
\title{An Analytic Study of the $\bf E\bigotimes e$ Jahn - Teller Polaron}
\author{Heinz Barentzen}
\address{Max--Planck--Institut f\"ur Festk\"orperforschung,
Heisenbergstra\ss e 1,\\
70569 Stuttgart, Federal Republic of Germany }
\maketitle
\widetext
\vspace{1cm}
\begin{abstract} 
\leftskip 54.8pt
\rightskip 54.8pt

An analytic study is presented of the $E\bigotimes e$ Jahn - Teller (JT) polaron, consisting 
of a mobile $e_g$ electron linearly coupled to the local $e_g$ normal vibrations of a periodic
array of octahedral complexes. Due to the linear coupling, the parity operator $\cal K$ and
the angular momentum operator $\cal J$ commute with the JT part and cause a twofold degeneracy 
of each JT eigenvalue. This degeneracy is lifted by the anisotropic hopping term. The Hamiltonian 
is then mapped onto a new Hilbert space, which is isomorphic to an eigenspace of $\cal J$ 
belonging to a fixed angular momentum eigenvalue $j>0$. In this representation, the Hamiltonian 
depends explicitly on $j$ and decomposes into a Holstein term and a residual JT interaction. 
While the ground state of the JT polaron is shown to belong to the sector $j=1/2$, the Holstein 
polaron is obtained for the ``unphysical'' value $j=0$. The new Hamiltonian is then subjected 
to a variational treatment, yielding the dispersion relations and effective masses for both 
kinds of polarons. The calculated polaron masses are in remarkably good agreement with recent 
quantum Monte Carlo data. The possible relevance of our results to the magnetoresistive manganite 
perovskites is briefly discussed.

\end{abstract}
\vspace{0.5cm}
\narrowtext
\widetext

\newpage
\subsection*{I.~INTRODUCTION}

The Jahn-Teller (JT) effect describes the interaction of lattice vibrational modes with
orbitally degenerate electronic states and thus refers to a particular type of
electron-phonon coupling.$^1$ Although this effect has proved indispensable for a proper
understanding of the physics of a variety of systems, ranging from paramagnetic ions in 
nonmagnetic crystals$^2$ to structural phase transitions,$^3$ it is fair to say that its
role in condensed-matter physics has been marginal for a long time.

For nearly a decade, however, the significance of the JT effect is undergoing a profound 
change, triggered by the discovery of superconductivity in the fullerides$^4$ and of very
large (``colossal'') magnetoresistance (CMR) in the manganite perovskites.$^5$ Because of 
their high symmetry, both classes of compounds fulfil the requirement for a JT interaction 
to occur, and numerous experiments seem to indicate that this is, in fact, the case. 
Manifestations of the JT effect in the fullerides have been reviewed by O'Brien and 
Chancey,$^6$ those in the manganites by Millis.$^7$ So far, however, there is no consensus 
as to the relative importance of the JT coupling in these materials. In their search of the 
origin of the CMR effect,$^5$ e.g., Millis {\em et al.}$^8$ argued that double exchange,$^9$ 
designed as a mechanism to induce ferromagnetic order in doped manganites, is not sufficient 
to account for the resistivity data and suggested that JT polaron formation is essential, 
whereas other authors$^{10}$ invoke ferromagnetic spin polarons to explain the effect. 
Problems of this kind could possibly be resolved by means of a detailed analytic theory of the 
JT polaron, yielding the (approximate) ground-state energy together with the corresponding 
eigenvector. Since the eigenstates of JT systems are {\em vibronic} in nature,$^1$ they may 
give rise to unexpected results for expectation values and correlation functions.

As a first step in this direction, we study the JT polaron of symmetry type $E\bigotimes e$,
which is most conveniently introduced by recalling some basic properties of 
La$_{1-x}$Ca$_x$MnO$_3$, a representative of the manganite family.$^7$ Each unit cell of the 
crystal contains an octahedral MnO$_6$ complex and an average number of 4-x $d$ electrons. 
Since the Hund's rule coupling is believed to be very strong, the spins of all the $d$ 
electrons are ferromagnetically aligned. Due to the crystal field produced by the oxygen 
ligands (point group O$_h$) the D state of the free Mn ion splits into a threefold degenerate 
$t_{2g}$ and a twofold degenerate $e_g$ level. Three of the electrons go into the tightly 
bound $t_{2g}$ orbitals forming a core spin of magnitude 3/2, while the remaining 1-x $d$ 
electrons occupy the $e_g$ orbitals and are mobile. To study the formation of the polaron we 
start with pure CaMnO$_3$ ($x=1$), where only the $t_{2g}$ orbitals are filled, and imagine 
that {\em one} additional electron is injected into the system (e.g., by replacing one Ca ion 
by La). The extra electron must go into the $e_g$ levels and, by virtue of symmetry, may couple 
to the $e_g$ normal vibrations of the octahedral complex. This type of vibronic interaction, 
where both the electron and the vibrational modes are of $e_g$ symmetry, is referred to as 
$E\bigotimes e$ JT coupling. In addition, the electron is allowed to move in a band composed 
of the local $e_g$ doublets. The resulting quasiparticle, consisting of the mobile $e_g$
electron and the concomitant $e_g$ distortion of the MnO$_6$ octahedra, is designated as 
$E\bigotimes e$ JT polaron. A somewhat simpler system, the $E\bigotimes b$ JT polaron, where 
$b$ denotes a non-degenerate representation of the tetragonal site group, has already been 
treated by H\"ock, Nickisch and Thomas$^{11}$ nearly two decades ago. 

In Sec. II we introduce our model, together with the angular momentum operator $\cal J$ and the 
parity operator $\cal K$. Since only linear JT coupling is considered, these operators commute 
with the JT Hamiltonian, but not with the (anisotropic) hopping term. In Sec. III we show that 
$\cal K$ generates new fermion operators such that both $\cal J$ and the JT term assume diagonal 
form with respect to the new fermionic basis. The spectrum of $\cal J$ is determined in Sec. IV, 
where we recover the well-known result that all eigenvalues of $\cal J$ are half-integral. We 
also show by rather general arguments that each eigenvalue of the JT Hamiltonian is still 
twofold degenerate. This degeneracy will be lifted by the hopping term. In Sec. V we construct
a representation of the original Hamiltonian on a new Hilbert space, which is isomorphic to an
eigenspace of $\cal J$ belonging to a fixed angular momentum eigenvalue $j>0$. The new
Hamiltonian depends explicitly on this quantum number and decomposes into a quasi-Holstein term
and a residual JT interaction. This is the optimal form, which can be reached by purely analytic
means, and elucidates the close relationship between the JT and the Holstein polaron. While the
latter is obtained for the ``unphysical'' value $j=0$, the ground state of the JT polaron is
shown to belong to the sector $j=1/2$. A variational treatment of the new Hamiltonian is outlined
in Sec. VI, where also some ground-state properties like the dispersion relations and effective 
masses for both kinds of polarons are presented. Our results are summarized in Sec. VII.

\subsection*{II.~THE MODEL HAMILTONIAN}

In the $E\bigotimes e$ JT polaron the state of the electron is completely specified by the
vectors $|i\gamma\rangle$, where $i$ denotes the cell index and $\gamma=x,z$ the components
of the $e_g$ doublet (spin indices are omitted since only a single $e_g$ electron is considered).
The wave functions $\langle\vec{r}|ix\rangle$ and $\langle\vec{r}|iz\rangle$ transform like 
the orbitals $d_{x^2-y^2}$ and $d_{3z^2-r^2}$, respectively, forming a local basis of the 
$E_g$ representation associated with each unit cell. The state of the electron may be more 
conveniently specified by the operators $e_{i\gamma}^{\dagger}$ and $e_{i\gamma}$, where 
$e_{i\gamma}^{\dagger}$ ($e_{i\gamma}$) creates (annihilates) an $e_g$ electron in the orbital 
state $\gamma=x$ or $z$ at lattice site $i$. Similarly, the $e_g$ distortions of the MnO$_6$ 
octahedra may be described either in terms of the local normal coordinates $Q_{i\gamma}$ or, 
more conveniently, by the bosonic creation and annihilation operators $a_{i\gamma}^{\dagger}$ 
and $a_{i\gamma}$, respectively, where the indices have the same meaning as above. The 
Hamiltonian used for the description of the $E\bigotimes e$ JT polaron reads

\begin{equation}
{\cal H} = {\cal H}_t + {\cal H}_v + {\cal H}_{JT} ,
\end{equation}
where

\begin{mathletters}
\begin{equation}
{\cal H}_t = -t\sum_{ia}\vec{e}_i{}^{\dagger}\cdot {\bf h}_a\cdot \vec{e}_{i+a}
\end{equation}
is the transfer or hopping term to be discussed below,

\begin{equation}
{\cal H}_v = \hbar\Omega\sum_i \vec{a}_i{}^{\dagger}\cdot\vec{a}_i
\end{equation}
describes the $e_g$ normal vibrations of frequency $\Omega$, while

\begin{equation}
{\cal H}_{JT} = g\hbar\Omega\sum_i \vec{e}_i{}^{\dagger}[(a_{ix}^{\dagger}+a_{ix})\sigma^x
-(a_{iz}^{\dagger}+a_{iz})\sigma^z]\vec{e}_i
\end{equation}
\end{mathletters}
represents the $E\bigotimes e$ JT coupling, where $\sigma^a$ ($a=x,y,z$) are the Pauli
matrices and the coupling strength is expressed by the dimensionless parameter $g$. The
column vectors 

\begin{equation}
\vec{e}_i = {e_{iz}\choose e_{ix}} \quad\mbox{and}\quad \vec{a}_i = {a_{iz}\choose a_{ix}} ,
\end{equation}
as well as their associated row vectors $\vec{e}_i{}^{\dagger}$ and $\vec{a}_i{}^{\dagger}$, 
have been introduced for convenience and to avoid an accumulation of indices.

The somewhat unusual form of the hopping term (2a), where the summation over $a$ runs over the 
six nearest neighbors of site $i$, originates from the orbital degeneracy of the electronic 
states. In orbitally degenerate systems the transfer of electrons between neighboring sites 
depends on the orientation of the orbitals and the direction of the transfer. The matrix 
elements $h_a^{\gamma\gamma'}$ of the matrices ${\bf h}_a$ are entirely determined by symmetry. 
Their numerical values along the three cubic axes are tabulated in Ref. 12, where the interatomic 
matrix element $V_{dd\sigma}$ is related to our hopping integral $t$. In the electronic basis 
defined by the vector $\vec{e}_i$ in Eqs. (3), the matrices ${\bf h}_a$ take the explicit form 

\begin{eqnarray}
{\bf h}_{\pm x} &=& (2\sigma^0-\sqrt{3}\sigma^x-\sigma^z)/4 ,\nonumber\\
{\bf h}_{\pm y} &=& (2\sigma^0+\sqrt{3}\sigma^x-\sigma^z)/4 , \\
{\bf h}_{\pm z} &=& (\sigma^0+\sigma^z)/2 ,\nonumber
\end{eqnarray}
where $\sigma^0$ denotes the $2\times 2$ unit matrix.

In Eq. (2c) we have restricted ourselves to linear JT coupling, adopted by the majority
of authors.$^{13,14}$ Moreover, our model does not contain the intersite coupling of the 
normal modes, which one intuitively expects since oxygens are shared between adjacent 
MnO$_6$ octahedra. These coupling terms will give rise to optical phonon branches and, as 
was recently pointed out by Hotta {\em et al.}$^{15}$ and Popovic and Satpathy,$^{16}$ to 
collective effects such as orbital ordering. Other terms like, e.g., the Hund's rule
coupling should also be included in a more rigorous treatment. Hence, the neglect of all 
these couplings might render our model somewhat unrealistic; we hope to include these terms 
in future work.

Since $\cal H$ is invariant under lattice translations, the total crystal momentum $\bf P=K+Q$ 
is a conserved quantity. Here the operators ${\bf K}=\sum_{\bf k}{\bf k}\,\vec{e}_{\bf k}{}^
{\dagger}\cdot\vec{e}_{\bf k}$ and ${\bf Q}=\sum_{\bf q}{\bf q}\,\vec{a}_{\bf q}{}^{\dagger}
\cdot\vec{a}_{\bf q}$ denote the crystal momenta of the electron and the $e_g$ vibrational
modes, respectively, where $\vec{e}_{\bf k}$ and $\vec{a}_{\bf q}$ are the Fourier transforms 
of $\vec{e}_i$ and $\vec{a}_i$. In addition to $\bf P$, there are operators which commute with 
${\cal H}_v$ and ${\cal H}_{JT}$, but not with the hopping term (unless ${\cal H}_t$ is assumed
to be isotropic). Although these operators are not strictly conserved, they often greatly 
facilitate the diagonalization of $\cal H$, as we shall see below. There are two operators of 
this kind which prove particularly useful:

\noindent
(i) the {\em parity operator}

\begin{mathletters}
\begin{equation}
\cal K = GR ,
\end{equation}
where

\begin{equation}
{\cal G} = \exp\left(i\pi\sum_i \vec{a}_i{}^{\dagger}\cdot\vec{a}_i\right)
\end{equation}
and

\begin{equation}
{\cal R} = \exp\left[i(\pi/2)\sum_i \vec{e}_i{}^{\dagger}\cdot(\sigma^y-\sigma^0)
\cdot\vec{e}_i\right] ;
\end{equation}
\end{mathletters}

\noindent
(ii) the {\em angular momentum operator}

\begin{mathletters}
\begin{equation}
{\cal J = M} - \frac{1}{2}\sum_i\vec{e}_i{}^{\dagger}\cdot\sigma^y\cdot\vec{e}_i ,
\end{equation}
where 

\begin{equation}
{\cal M} = \sum_i\vec{a}_i{}^{\dagger}\cdot\sigma^y\cdot\vec{a}_i
\end{equation}
\end{mathletters}\noindent
is referred to as vibrational angular momentum. The spectral properties of these operators
will be discussed in the following sections. Here it suffices to mention that quantities 
similar to $\cal K$ and $\cal J$ also play an important role in isolated $E\bigotimes e$ JT 
centers, provided the JT coupling is {\em linear}.$^{17}$ In such systems the only eigenvalues 
of $\cal K$ are $1$ and $-1$, whereas those of $\cal J$ range over all half-odd integers.
The most important properties of the operator $\cal K$ may be summarized by the equations

\begin{equation}
{\cal K}^{\dagger}\vec{e}_i{\cal K} = \sigma^y\cdot\vec{e}_i , 
\quad {\cal K}^{\dagger}\vec{a}_i{\cal K} = - \vec{a}_i ,
\end{equation}
whose derivation rests on the well-known commutator expansion

\begin{displaymath}
e^S A e^{-S} = A + [S,A] + (2!)^{-1}[S,[S,A]] + \cdots .
\end{displaymath}
Using (7) we see that both ${\cal H}_v+{\cal H}_{JT}$ and $\cal J$ are left invariant by 
$\cal K$, i.e., ${\cal H}_v+{\cal H}_{JT}$, $\cal K$ and $\cal J$ form a complete set of 
commuting operators:
\begin{equation}
[{\cal H}_v+{\cal H}_{JT},{\cal K}]=[{\cal H}_v+{\cal H}_{JT},{\cal J}]
=[{\cal K,J}]=0 .
\end{equation}

\subsection*{III.~ GENERATION OF NEW FERMION OPERATORS}

In this section we shall exploit the properties of the parity operator $\cal K$ to generate 
new fermionic creation and annihilation operators such that ${\cal H}_{JT}$ and $\cal J$ take 
diagonal form with respect to these operators. To this end we need another property of 
$\cal K$, which reads

\begin{equation}
{\cal K}^2 = 1 
\end{equation} 
and readily follows from Eqs. (7) and the relation $(\sigma^a)^2=\sigma^0$ valid for all 
Pauli matrices. Hence, as in an isolated JT center, the only eigenvalues of $\cal K$ are 
$\kappa =\pm 1$. Moreover, since $\cal K$ is also unitary, we have the additional relations 
${\cal K}={\cal K}^{-1}={\cal K}^{\dagger}$.

As the next step, we need the projection operator ${\cal P}_{\kappa}$ for selecting the 
subspace associated with the eigenvalue $\kappa$ of $\cal K$. According to L\"owdin,$^{18}$ 
${\cal P}_{\kappa}$ is given by the expression

\begin{equation}
{\cal P}_{\kappa} = \frac{1}{2}\left(1 + \kappa{\cal K}\right) \quad (\kappa = \pm 1) ,
\end{equation} 
which, apart from being Hermitian, has the properties

\begin{mathletters}
\begin{eqnarray}
\sum_{\kappa}{\cal P}_{\kappa} & = & 1 ,\\
{\cal P}_{\kappa}{\cal P}_{\kappa'} & = & \delta_{\kappa\kappa'}{\cal P}_{\kappa} .
\end{eqnarray}
\end{mathletters}\noindent
We also have the obvious relation ${\cal KP}_{\kappa}=\kappa{\cal P}_{\kappa}$, implying that 
the subspace projected out by ${\cal P}_{\kappa}$ is an eigenspace of $\cal K$ to the 
eigenvalue $\kappa$. Property (11a) allows us to decompose the Hamiltonian and the angular 
momentum operator $\cal J$ into components acting on the eigenspaces of $\cal K$ as follows

\begin{mathletters}
\begin{equation}
{\cal H}=\sum_{\kappa\kappa'}{\cal P}_{\kappa}{\cal H}_t{\cal P}_{\kappa'} + {\cal H}_v + 
\sum_{\kappa}{\cal P}_{\kappa}{\cal H}_{JT}{\cal P}_{\kappa} ,
\end{equation}
\begin{equation}
{\cal J} = \sum_{\kappa}{\cal P}_{\kappa}{\cal J}{\cal P}_{\kappa} ,
\end{equation}
\end{mathletters}\noindent
where we have used Eq. (11b) and the fact that ${\cal H}_v$, ${\cal H}_{JT}$, and $\cal J$ 
commute with ${\cal P}_{\kappa}$. Since the hopping term does not commute with $\cal K$, the 
eigenspaces of the latter are mixed by ${\cal H}_t$, as was to be expected.

To obtain Eqs. (12) in explicit form, we need to calculate the operators $\vec{e}_i
{\cal P}_{\kappa}$. Using (7) and (10) we find

\begin{equation}
\vec{e}_i{\cal P}_{\kappa}=\frac{1}{2}(\sigma^0+\kappa\sigma^y{\cal K})\cdot\vec{e}_i
=\frac{1}{2}{e_{iz}-i\kappa{\cal K}e_{ix}\choose e_{ix}+i\kappa{\cal K}e_{iz}} ,
\end{equation}
and we shall now prove that the products ${\cal K}e_{i\gamma}$ ($\gamma=x,z$) on the right side 
of Eq. (13) may be replaced by ${\cal G}e_{i\gamma}$, where $\cal G$ is defined by Eq. (5b). To 
show this, let $|\Psi\rangle$ be an arbitrary vector of the underlying {\em single-particle} 
Hilbert space,

\begin{equation}
|\Psi\rangle = \sum_{i\gamma} \Psi_{i\gamma} e_{i\gamma}^{\dagger}|0\rangle ,
\end{equation}
where $\Psi_{i\gamma}$ are pure functions of the Bose operators $a_{i\gamma}$, $a_{i\gamma}
^{\dagger}$ and $|0\rangle$ denotes the common vacuum for all particles. If ${\cal K}e_{i\gamma}$ 
is now applied to $|\Psi\rangle$ and use is made of the fact that $\cal R$ of Eq. (5c) commutes 
with $\Psi_{i\gamma}$ ($\gamma=x,z$), the result is

\begin{displaymath}
{\cal K}e_{i\gamma}|\Psi\rangle ={\cal K}\Psi_{i\gamma}|0\rangle ={\cal G}\Psi_{i\gamma}|0\rangle
={\cal G}e_{i\gamma}|\Psi\rangle .
\end{displaymath}
Hence, ${\cal K}e_{i\gamma}={\cal G}e_{i\gamma}$ on the entire Hilbert space, which proves our 
claim. Expression (13) may thus be rewritten as 

\begin{mathletters}
\begin{equation}
\vec{e}_i{\cal P}_{\kappa} = \frac{1}{2}{e_{iz}-i\kappa{\cal G}e_{ix}\choose e_{ix}+i\kappa 
{\cal G}e_{iz}} = \vec{u}_{\kappa}d_{i\kappa} ,
\end{equation}
where

\begin{equation}
\vec{u}_{\kappa} = \frac{1}{\sqrt{2}} {1 \choose i\kappa{\cal G}}
\end{equation}
is a normalized vector (i.e., $\vec{u}_{\kappa}{}^{\dagger}\cdot\vec{u}_{\kappa}=1$), while the 
quantities

\begin{equation}
d_{i\kappa} = \frac{1}{\sqrt{2}}(e_{iz}-i\kappa{\cal G}e_{ix}) 
\end{equation}
\end{mathletters}\noindent
behave like ordinary fermion operators, i.e.,

\begin{equation}
[d_{i\kappa}, d^{\,\dagger}_{j\kappa'}]_+ = \delta_{ij}\delta_{\kappa\kappa'}, \quad
[d_{i\kappa}, d_{j\kappa'}]_+ = 0 .
\end{equation}
However, due to the presence of the operator $\cal G$ in Eq. (15c), the $d_{i\kappa}$, 
$d_{i\kappa}^{\,\dagger}$ {\em cease to commute} with the Bose operators $a_{i\gamma}$ and 
$a_{i\gamma}^{\dagger}$, but continue to commute with quadratic forms like $\cal G$ and the 
vibrational angular momentum $\cal M$ of Eq. (6b). 
    
With the help of Eqs. (15), the various parts of the Hamiltonian (12a) may now be expressed in 
terms of the new fermion operators $d_{i\kappa}$ and $d_{i\kappa}^{\,\dagger}$ ($\kappa=\pm 1$). 
We start with the hopping term ${\cal H}_t$, which is transformed into

\begin{mathletters}
\begin{equation}
{\cal H}_t = -t\sum_{ia}\sum_{\kappa\kappa'}d_{i\kappa}^{\,\dagger}\tau_a^{\kappa\kappa'}
d_{i+a,\kappa'} ,
\end{equation}
\begin{equation}
\tau_a^{\kappa\kappa'} = \vec{u}_{\kappa}{}^{\dagger}\cdot{\bf h}_a\cdot\vec{u}_{\kappa'} ,
\end{equation}
\end{mathletters}\noindent
where the matrices ${\bf h}_a$ are given by Eqs. (4). In the basis defined by the vector 
$\vec{d}_i={d_{i+}\choose d_{i-}}$, Eqs. (17) may also be written as

\begin{mathletters}
\begin{equation}
{\cal H}_t = -t\sum_{ia}\vec{d}_i{}^{\dagger}\cdot\mbox{\boldmath$\tau$}_a\cdot\vec{d}_{i+a} ,
\end{equation}
where the new hopping matrices $\mbox{\boldmath$\tau$}_a$ read  

\begin{eqnarray}
\mbox{\boldmath$\tau$}_{\pm x} &=& (2\sigma^0-\sigma^x-\sqrt{3}{\cal G}\sigma^y)/4 , \nonumber\\
\mbox{\boldmath$\tau$}_{\pm y} &=& (2\sigma^0-\sigma^x+\sqrt{3}{\cal G}\sigma^y)/4 , \\
\mbox{\boldmath$\tau$}_{\pm z} &=& (\sigma^0+\sigma^x)/2 , \nonumber
\end{eqnarray}
\end{mathletters}\noindent
being now explicit functions of the operator $\cal G$. The vibrational part ${\cal H}_v$ 
remains unchanged, while the JT coupling takes the form

\begin{equation}
{\cal H}_{JT} = g\hbar\Omega\sum_{i\kappa}d_{i\kappa}^{\,\dagger}[i\kappa(a_{ix}^{\dagger}
+a_{ix}){\cal G} - (a_{iz}^{\dagger}+a_{iz})]d_{i\kappa} .
\end{equation}
In deriving this result we have used the relations $a_{i\gamma}\vec{u}_{\kappa}=\vec{u}_{-\kappa}
a_{i\gamma}$, $\vec{u}_{\kappa}{}^{\dagger}\cdot\sigma^x\cdot\vec{u}_{-\kappa}=-i\kappa{\cal G}$, 
and $\vec{u}_{\kappa}{}^{\dagger}\cdot\sigma^z\cdot\vec{u}_{-\kappa}=1$. Finally, the angular 
momentum operator $\cal J$ of Eq. (12b) is obtained as

\begin{mathletters}
\begin{equation}
{\cal J = M} - ({\cal G}/2)\sum_{i\kappa}\kappa d_{i\kappa}^{\,\dagger}d_{i\kappa} ,  
\end{equation}
and we see that ${\cal H}_{JT}$ and $\cal J$ are now {\em diagonal} with respect to the new fermion 
operators. The vibrational part of $\cal J$ is, however, still nondiagonal. The diagonalization 
of $\cal M$ is readily accomplished by means of the substitutions

\begin{eqnarray*}
a_{iz} &\to & \frac{1}{\sqrt{2}}(a_{iz}+a_{ix}) ,\\
a_{ix} &\to & \frac{-i}{\sqrt{2}}(a_{iz}-a_{ix}) ,
\end{eqnarray*}
which leave ${\cal H}_t$ and ${\cal H}_v$ invariant, while $\cal M$ is brought to the diagonal 
form

\begin{equation}
{\cal M} = \sum_i (a_{ix}^{\dagger}a_{ix}-a_{iz}^{\dagger}a_{iz}) =\sum_i{\cal M}_i .
\end{equation}
\end{mathletters}\noindent
The JT coupling is transformed into the expression

\begin{equation}
{\cal H}_{JT} = -\sqrt{2}g\hbar\Omega\sum_{i\kappa}d_{i\kappa}^{\,\dagger}(a_{ix}\Pi_{\kappa}+ 
a_{iz}\Pi_{-\kappa} + \mbox{H.c.})d_{i\kappa} ,
\end{equation}
where the new projection operators

\begin{equation}
\Pi_{\kappa} = \frac{1}{2}(1 + \kappa{\cal G}) \quad (\kappa=\pm 1) 
\end{equation}
have been introduced. For later purposes we need the properties

\begin{mathletters}
\begin{eqnarray}
\Pi_{\kappa}a_{i\gamma} &=& a_{i\gamma}\Pi_{-\kappa} ,\\
\Pi_{\kappa}\Pi_{\kappa'} &=& \delta_{\kappa\kappa'}\Pi_{\kappa} ,
\end{eqnarray}
\begin{eqnarray}
\Pi_{\kappa}+\Pi_{-\kappa} &=& 1 ,\\
\Pi_{\kappa}-\Pi_{-\kappa} &=& \kappa{\cal G} ,
\end{eqnarray}
\end{mathletters}\noindent
which follow immediately from definition (22). Before we set out to develop strategies for 
dealing with the complicated vibrational terms in ${\cal H}_{JT}$, we shall first derive the 
spectrum of $\cal J$ and investigate the possible symmetries and degeneracies of the JT 
Hamiltonian.

\subsection*{IV.~ SYMMETRIES AND DEGENERACIES}

In isolated JT centers there often exist ``hidden'' symmetries giving rise to unexpected 
degeneracies of the energy levels. In the linearly coupled $E\bigotimes e$ JT center, e.g., 
all eigenvalues are twofold degenerate, and higher-order coupling terms are necessary to 
(partially) remove the degeneracy.$^1$ This section is devoted to a study of these 
degeneracies in JT crystals like the manganites.

To investigate the possible symmetries of our system, we first need the spectrum of the 
angular momentum operator $\cal J$, Eqs. (20). Since $\cal J$, $\cal K$, and ${\cal H}_v$ 
commute with each other, these operators possess common eigenstates. The latter are of the 
form

\begin{mathletters}
\begin{equation}
|\Psi_{nj\kappa}^{\,0}\rangle = \sum_i C_{i\kappa}d_{i\kappa}^{\,\dagger}\prod_i|n_i,m_i\rangle  
\quad (\kappa=\pm 1) ,
\end{equation}
where $C_{i\kappa}$ are coefficients and the product extends over the local eigenstates of 
the isotropic oscillator in two dimensions:$^{19}$

\begin{equation}
|n_i,m_i\rangle = \frac{(a_{ix}^{\dagger})^{(n_i+m_i)/2}(a_{iz}^{\dagger})^
{(n_i-m_i)/2}}{\sqrt{\left(\frac{n_i+m_i}{2}\right)!\left(\frac{n_i-m_i}{2}
\right)!}}|0\rangle ,
\end{equation}
\begin{equation}
n_i = 0,1,2,\cdots ; \: m_i = n_i,n_i-2,\cdots,-n_i .
\end{equation}
\end{mathletters}\noindent
In fact, a simple calculation shows that $|\Psi_{nj\kappa}^{\,0}\rangle$ is an eigenstate of 
$\cal K$, $\cal J$, and ${\cal H}_v$ with the respective eigenvalues $\kappa$, $j_{\kappa}$, 
and $E_{nj\kappa}^{\,0}$, where 

\begin{mathletters}
\begin{equation}
E_{nj\kappa}^{\,0} = n\hbar\Omega ,\quad j_{\kappa}= m - \frac{\kappa}{2}(-1)^n,
\end{equation}
\begin{equation}
n = \sum_i n_i , \quad m = \sum_i m_i .
\end{equation}
\end{mathletters}\noindent
There are many other linearly independent eigenvectors of $\cal K$, $\cal J$, and 
${\cal H}_v$ belonging to the same eigenvalues: all vectors of the form (24a), whose quantum 
numbers $n_i$ and $m_i$ satisfy the constraints (24c) and (25b), are also eigenstates with 
the required properties. Together they span a vector space ${\cal U}_j^0$, and we see that 
the energies $E_{nj\kappa}^{\,0}$ are highly degenerate.$^{20}$  

For a single site we have the relation $m_i=n_i-2p_i$ ($p_i=0,1,\cdots,n_i$) which, after
summation over all cells, becomes $m=n-2p$ ($p=0,1,\cdots,n$). Thus, for a given $n$, the
quantum number $m$ may take the $n+1$ integral values $m=n,n-2,\cdots,-n$, whence we conclude 
that the angular momentum quantum numbers $j_{\kappa}$ must all be {\em half-integral}, as 
in an isolated $E\bigotimes e$ JT center with linear coupling.$^1$ A more detailed analysis 
of $j_{\kappa}$, Eq. (25a), requires a distinction between even and odd $n$ ($n$ and $m$ 
always have the same parity, both being either even or odd). It is not difficult to verify that, 
for fixed $\kappa$, both cases yield the same eigenvalues $j_{\kappa}$ so that we may restrict 
ourselves to even $m$ ($n$). The angular momentum quantum numbers may thus also be written as
 
\begin{displaymath}
j_{\kappa} = m - \kappa/2  \quad (m=0,\pm 2,\pm 4,\cdots)
\end{displaymath}
or, explicitly:

\begin{equation}
j_{\kappa}=\left\{ \begin{array}{l@{\hspace{0.5cm}}l}
\cdots,-5/2,-1/2,3/2,7/2,\cdots  & \kappa=1 \\
\cdots,-7/2,-3/2,1/2,5/2,\cdots  & \kappa=-1 .
\end{array} \right.
\end{equation} 

We now set out to examine more closely the structure of $\cal J$ and ${\cal H}_{JT}$ which, 
for the present purpose, are written as ${\cal J}=\sum_{\kappa}{\cal J}^{(\kappa)}$ and 
${\cal H}_{JT}=\sum_{\kappa}{\cal H}_{JT}^{(\kappa)}$. A glance at Eqs. (20) and (21) then 
reveals that the substitution $d_{i\kappa}\to d_{i,-\kappa}$, combined with the interchange 
$a_{ix}\leftrightarrow a_{iz}$ (these operations correspond to canonical transformations and, 
hence, do not affect the eigenvalues), has the effect that ${\cal H}_{JT}^{(\kappa)}\to 
{\cal H}_{JT}^{(-\kappa)}$ and ${\cal J}^{(\kappa)}\to -{\cal J}^{(-\kappa)}$. Denoting the 
eigenvalues of ${\cal H}_{JT}^{(\kappa)}$ and ${\cal J}^{(\kappa)}$ by $E_{nj\kappa}$ and 
$j_{\kappa}$, respectively, and the common eigenvectors of these operators by 
$|\Psi_{nj\kappa}\rangle$, we may thus draw the following conclusions:

\noindent
1. ${\cal H}_{JT}^{(+)}$ and ${\cal H}_{JT}^{(-)}$ have the same eigenvalues which must, 
therefore, be independent of $\kappa$: $E_{nj\kappa}=E_{nj}$. Since the corresponding 
eigenvectors $|\Psi_{nj+}\rangle$ and $|\Psi_{nj-}\rangle$ are orthogonal by virtue of Eq. 
(11b), each eigenvalue $E_{nj}$ of ${\cal H}_{JT}$ is necessarily {\em twofold degenerate}.

\noindent
2. The spectra of ${\cal J}^{(+)}$ and ${\cal J}^{(-)}$ have the property that to any 
positive eigenvalue $j_{+}$ of ${\cal J}^{(+)}$ there is always a negative eigenvalue $j_{-} 
= -j_{+}$ of ${\cal J}^{(-)}$ and {\em vice versa}. This property is most clearly reflected 
by Eq. (26). Since the eigenvalues of ${\cal H}_{JT}$ are independent of $\kappa$, they can 
only depend on $j\equiv |j_{\kappa}|=1/2,3/2,5/2,\cdots$.  

\noindent
Thus, we see that the huge degeneracy of the eigenvalues $E_{nj\kappa}^{\,0}$ of ${\cal H}_v$ 
(see Ref. 20) is nearly completely lifted by the JT Hamiltonian. The remaining twofold 
degeneracy of ${\cal H}_{JT}$, which is of the same origin as that in the linearly coupled 
$E\bigotimes e$ JT center, will be removed by the hopping term ${\cal H}_t$ (apart from
accidental degeneracy). 
 
Only relatively few of the vectors contained in ${\cal U}_j^0$ are simultaneous eigenstates 
of $\cal J$ and ${\cal H}_{JT}$. To find these eigenvectors, we shall take advantage of the 
existence of a simple operator $\cal C$, which also commutes with $\cal J$ and is of help to 
select the proper candidates. This operator will be shown in Sec. V to emerge from the JT 
Hamiltonian and reads

\begin{equation}
{\cal C}=\sum_{i\kappa}d_{i\kappa}^{\,\dagger}(1-\kappa{\cal GM}_i)d_{i\kappa} .
\end{equation}
There are, in fact, {\em two} orthogonal sets of eigenstates both belonging to the {\em same} 
quantum number $j$, but to {\em different} eigenvalues of $\cal C$. For {\em positive} $j$ the 
two sets are represented by the vectors (similar vectors have been constructed in Ref. 21) 

\begin{mathletters}
\begin{eqnarray}
|\Psi_{nj\kappa}^+\rangle &=& \sum_i\sum_{n_i=0}^{\infty}C_{i\kappa}^+(n_i)d_{i\kappa}^{\,
\dagger}|m+2n_i,m\rangle\prod_{l\neq i}|0_l,0_l\rangle ,\\
|\Psi_{nj\kappa}^-\rangle &=& \sum_i\sum_{n_i=0}^{\infty}C_{i\kappa}^-(n_i)d_{i\kappa}^{\,
\dagger}|m+2n_i+1,m+1\rangle\prod_{l\neq i}|0_l,0_l\rangle ,
\end{eqnarray}
where $m$ is the {\em same} for all $i$ and may assume the values

\begin{equation}
m = j-1/2 = 0,1,2,\cdots .
\end{equation}
\end{mathletters}\noindent
It is then straightforward to verify the eigenvalue equations

\begin{mathletters}
\begin{eqnarray}
{\cal J}|\Psi_{nj\kappa}^{\pm}\rangle &=& j|\Psi_{nj\kappa}^{\pm}\rangle ,\\
{\cal C}|\Psi_{nj\kappa}^{\pm}\rangle &=& (1/2\pm j)|\Psi_{nj\kappa}^{\pm}\rangle ,
\end{eqnarray}
whose validity requires that

\begin{equation}
\kappa = -(-1)^m = (-1)^{j+1/2} .
\end{equation}
\end{mathletters}\noindent
Common eigenstates of $\cal J$ and $\cal C$ for negative $j$ also exist, but are not needed here.

Hence, the vectors $|\Psi_{nj\kappa}^{\pm}\rangle$ are common eigenstates of $\cal J$ and 
$\cal C$. Physically they represent {\em polaronic} states, where the electron is accompanied by 
an on-site distortion (vibrational excitation) of the molecular complex, all other complexes 
not coinciding with the location of the electron being left in their vibrational ground states. 
All vectors (28a), where $m$ is given and $\kappa$ is fixed by Eq. (29c), form a subspace
${\cal U}_j^+$, while those obtained from $|\Psi_{nj\kappa}^-\rangle$ span a subspace 
${\cal U}_j^-$, which is orthogonal to ${\cal U}_j^+$. The direct sum of these spaces will be
denoted as ${\cal U}_j$ to remind us that this is an eigenspace of $\cal J$ to the eigenvalue 
$j = m+1/2 > 0$. Since this is the most general eigenspace, which is compatible with the
existence of the operator $\cal C$, the simultaneous eigenvectors $|\Psi_{nj\kappa}\rangle$ of 
${\cal H}_{JT}$ and $\cal J$ are necessarily all contained in ${\cal U}_j$.

\subsection*{V.~ REPRESENTATION OF $\mbox{\boldmath$\cal H$}$ FOR FIXED QUANTUM NUMBER 
$\mbox{\boldmath$j$}$}

Our main goal in this section is to construct a {\em representation} of the Hamiltonian on a 
subspace ${\cal V}_j$, which is defined to be isomorphic to the space ${\cal U}_j$ introduced at 
the end of the preceding section. We start with the JT term, whose representation rests on the 
operators

\begin{mathletters}
\begin{eqnarray}
{\cal A} &=& \sqrt{2}\sum_{i\kappa}d_{i\kappa}^{\,\dagger}(a_{ix}\Pi_{\kappa}
+ a_{iz}\Pi_{-\kappa})d_{i\kappa} ,\\
{\cal A}^{\dagger} &=& \sqrt{2}\sum_{i\kappa}d_{i\kappa}^{\,\dagger}(a_{ix}^
{\dagger}\Pi_{-\kappa} + a_{iz}^{\dagger}\Pi_{\kappa})d_{i\kappa} ,
\end{eqnarray}
\end{mathletters}\noindent
allowing the JT Hamiltonian (21) to be written in the simple form

\begin{equation}
{\cal H}_{JT} = - g\hbar\Omega ({\cal A} + {\cal A}^{\dagger}) .
\end{equation}
The merits of this seemingly trivial reformulation will become obvious later on. As the next 
step, we calculate the products ${\cal AA}^{\dagger}$ and ${\cal A}^{\dagger}{\cal A}$. Using 
properties (23a) and (23b) we find the expressions

\begin{mathletters}
\begin{eqnarray}
{\cal AA}^{\dagger} &=& 2\sum_{i\kappa} d_{i\kappa}^{\,\dagger}(1+a_{ix}^{\dagger}a_{ix}
\Pi_{-\kappa}+a_{iz}^{\dagger}a_{iz}\Pi_{\kappa})d_{i\kappa} ,\\
{\cal A}^{\dagger}{\cal A} &=& 2\sum_{i\kappa} d_{i\kappa}^{\,\dagger}(a_{ix}^{\dagger}
a_{ix}\Pi_{\kappa}+a_{iz}^{\dagger}a_{iz}\Pi_{-\kappa})d_{i\kappa} ,
\end{eqnarray}
\end{mathletters}\noindent
which enable us to set up the commutator $[{\cal A,A}^{\dagger}\,]$ and the anticommutator 
$[{\cal A,A}^{\dagger}\,]_+$. Using properties (23c) and (23d) we obtain

\begin{mathletters}
\begin{eqnarray}
\frac{1}{2}[{\cal A,A}^{\dagger}\,] &=& \sum_{i\kappa}d_{i\kappa}^{\,\dagger}
(1-\kappa{\cal G M}_i)d_{i\kappa}\equiv {\cal C} ,\\
\frac{1}{2}[{\cal A,A}^{\dagger}\,]_+ &=& \sum_{i\kappa}d_{i\kappa}^{\,\dagger}
(1+\vec{a}_i{}^{\dagger}\cdot\vec{a}_i)d_{i\kappa}\equiv {\cal N} ,
\end{eqnarray}
\end{mathletters}\noindent
where ${\cal C}=\frac{1}{2}[{\cal A,A}^{\dagger}\,]$ is the operator (27), whose eigenvalue 
problem is given by Eq. (29b). The result may be restated as follows: on the subspace 
${\cal U}_j^+$ the operator $\cal C$ reduces to a positive integer and takes the form

\begin{equation}
\frac{1}{2}[{\cal A,A}^{\dagger}\,]={\cal C}= m+1 = j+1/2 \quad (m\geq 0) ,
\end{equation}
whereas on ${\cal U}_j^-$ it reduces to the number $1/2-j$, which is negative for all $j > 1/2$. 
To appreciate this result, we now investigate the anticommutator $\cal N$, Eq. (33b). First of 
all one realizes that $\cal N$ is a {\em positive} operator. This property, together with the 
identity

\begin{mathletters}
\begin{equation}
{\cal N}={\cal A}^{\dagger}{\cal A}+{\cal C} ,
\end{equation}
imposes a constraint on $\cal C$ requiring that this must also be a positive operator. The only
way to guarantee that $\cal C$ and, hence, $\cal N$ are positive for all $j \geq 1/2$ is to
restrict the commutator to the subspace ${\cal U}_j^+$ as in Eq. (34). 

Particularly interesting are the commutators $[{\cal N,A}\,]$ and $[{\cal N,A}^{\dagger}\,]$, 
since they agree with those of ordinary Bose operators.$^{22}$ In fact, using Eqs. (30) and (33b) 
one obtains  

\begin{equation}
[{\cal N,A}\,]=-{\cal A} , \quad [{\cal N,A}^{\dagger}\,]={\cal A}^{\dagger}.
\end{equation}
\end{mathletters}\noindent
We also need to express ${\cal H}_v$ in terms of the operators $\cal A$ and ${\cal A}^{\dagger}$. 
To this end we calculate the expectation value of $\cal N$ in the state $|i\kappa\rangle = 
d_{i\kappa}^{\,\dagger}|0\rangle_e$, where $|0\rangle_e$ denotes the {\em electronic} vacuum. The 
result is $\langle i\kappa|{\cal N}|i\kappa\rangle=1+\vec{a}_i{}^{\dagger}\cdot\vec{a}_i$ which, 
after summation over all sites, yields

\begin{equation}
{\cal H}_v = \hbar\Omega\sum_i(\langle i\kappa|{\cal N}|i\kappa\rangle - 1) .
\end{equation}

We are now in a position to construct the desired representation of the JT Hamiltonian, as 
defined at the beginning of this section. Our idea is to map the Hilbert space ${\cal U}_j$ onto 
a new space ${\cal V}_j$, isomorphic to ${\cal U}_j$, and to construct operators on the new space 
satisfying the same relations as $\cal A$ and ${\cal A}^{\dagger}$ on ${\cal U}_j$ (this procedure 
is closely related to that employed in the bosonization of spin operators). The new space 
${\cal V}_j$ is spanned by all vectors of the form

\begin{mathletters}
\begin{eqnarray}
|\Phi_{nj\kappa}^+) &=& \sum_i\sum_{n_i=0}^{\infty}C_{i\kappa}^+(n_i)c_{i\kappa}^{\dagger}
\frac{(b_i^{\,\dagger})^{2n_i}}{\sqrt{(2n_i)!}}|0) ,\\
|\Phi_{nj\kappa}^-) &=& \sum_i\sum_{n_i=0}^{\infty}C_{i\kappa}^-(n_i)c_{i\kappa}^{\dagger}
\frac{(b_i^{\,\dagger})^{2n_i+1}}{\sqrt{(2n_i+1)!}}|0) ,
\end{eqnarray}
\end{mathletters}\noindent
where $b_i^{\,\dagger}$ and $c_{i\kappa}^{\dagger}$ create new bosons and fermions, respectively, 
and $|0)$ is the common vacuum of the new particles. The new fermion operators are similarly 
defined as our previous $d_{i\kappa}$ of Eq. (15c):

\begin{mathletters}
\begin{eqnarray}
c_{i\kappa} &=& \frac{1}{\sqrt{2}}(e_{iz}-i\kappa G e_{ix}) ,\\
G &=& \exp \left(i\pi\sum_i b_i^{\,\dagger}b_i\right) .
\end{eqnarray}
\end{mathletters}\noindent
The required isomorphism between ${\cal U}_j$ and ${\cal V}_j$ is achieved by the mapping
prescription

\begin{equation}
|\Psi_{nj\kappa}^{\pm}\rangle \leftrightarrow |\Phi_{nj\kappa}^{\pm}) ,
\end{equation}
which obviously establishes a one-to-one correspondence between all vectors of ${\cal U}_j$ and 
those of ${\cal V}_j$. As in our construction of the space ${\cal U}_j$, the new space may also be 
conceived as the direct sum of two orthogonal subspaces, ${\cal V}_j^+$ and ${\cal V}_j^-$, spanned 
by all vectors of the form (37a) and (37b), respectively. The quantum number $\kappa$ in Eqs. (37) 
may be assigned the value (29c), but we shall see below that the results are independent of this 
choice. 

On the new space ${\cal V}_j$ we now define the operator

\begin{mathletters}
\begin{equation}
A_j = \sum_{i\kappa}c_{i\kappa}^{\dagger}A_i^{(j)}b_ic_{i\kappa} ,
\end{equation}
where $A_i^{(j)}$ is self-adjoint and {\em explicitly} dependent on the quantum number $j$:

\begin{equation}
A_i^{(j)} = P_- + \left(1+\frac{2j}{b_i^{\,\dagger}b_i+1}\right)^{1/2}P_+ , 
\end{equation}
\begin{equation}
P_{\pm}=\frac{1}{2}(1\pm G).
\end{equation}
\end{mathletters}\noindent
The operators $P_{\pm}$, the analogues of our previous $\Pi_{\kappa}$, satisfy a set of relations 
differing from Eqs. (23) only in notation (we have, e.g., that $P_{\pm}b_i=b_iP_{\mp}$). We now set 
out to prove that $A_j$ and $A_j^{\dagger}$ satisfy the same algebraic relations on ${\cal V}_j$ as 
$\cal A$ and ${\cal A}^{\dagger}$ on the space ${\cal U}_j$ (another proof of the equivalence of
the operators $\cal A$ and $A_j$ in terms of their matrix elements is given in Appendix A). To this 
end we start by calculating the products $A_jA_j^{\dagger}$ and $A_j^{\dagger}A_j$. Using the 
well-known relation $F(b_i^{\,\dagger}b_i)b_i=b_iF(b_i^{\,\dagger}b_i-1)$, we find that

\begin{mathletters}
\begin{eqnarray}
A_jA_j^{\dagger} &=& \sum_{i\kappa}c_{i\kappa}^{\dagger}(b_i^{\,\dagger}b_i+2jP_+ +1)c_{i\kappa} ,\\
A_j^{\dagger}A_j &=& \sum_{i\kappa}c_{i\kappa}^{\dagger}(b_i^{\,\dagger}b_i + 2jP_-)c_{i\kappa} .
\end{eqnarray}
\end{mathletters}\noindent
By means of these relations, we may now evaluate the commutator $[A_j,A_j^{\dagger}\,]$ and the 
anticommutator $[A_j,A_j^{\dagger}\,]_+$. A simple calculation gives

\begin{mathletters}
\begin{eqnarray}
\frac{1}{2}[A_j,A_j^{\dagger}\,] &=& \sum_{i\kappa}c_{i\kappa}^{\dagger}(1/2 + jG)
c_{i\kappa}\equiv C_j ,\\
\frac{1}{2}[A_j,A_j^{\dagger}\,]_+ &=& \sum_{i\kappa}c_{i\kappa}^{\dagger}
(b_i^{\,\dagger}b_i+j+1/2)c_{i\kappa}\equiv N_j ,
\end{eqnarray}
\end{mathletters}\noindent
where $C_j$ and $N_j$ denote the analogues of our previous operators $\cal C$ and $\cal N$. 

If our formalism is to be meaningful, we expect the commutators $[N_j,A_j\,]$ and 
$[N_j,A_j^{\dagger}\,]$ to be the same as in Eqs. (35b), differing from the latter only in 
notation. This is in fact the case, for a straightforward calculation based on Eqs. (40a) and 
(42b) shows that

\begin{equation}
[N_j,A_j\,] = - A_j ,\quad [N_j,A_j^{\dagger}\,] = A_j^{\dagger} ,
\end{equation}
and we recover the Bose-like commutation relations (35b).$^{22}$ It still remains to be verified 
that the commutator (42a) agrees with that of Eq. (34). This is readily shown and follows from 
the observation that on ${\cal V}_j^+$ the operator $G$ acts like the unit operator, while on
${\cal V}_j^-$ it has the eigenvalue $-1$. Moreover, since $N_j$ is again a {\em positive}
operator, $C_j$ must also be positive by the same arguments as those used in conjunction with 
Eq. (34). Hence, on the allowed subspace ${\cal V}_j^+$ the commutator reduces to the positive 
integer

\begin{equation}
\frac{1}{2}[A_j,A_j^{\dagger}\,] = C_j = j+1/2 ,
\end{equation}
exactly like $\cal C$ on ${\cal U}_j^+$ (see Eq. (34)). Thus, we have shown that the operators 
$A_j,\:A_j^{\dagger},\:N_j$ and ${\cal A},\:{\cal A}^{\dagger},\:{\cal N}$ are defined on 
isomorphic Hilbert spaces and satisfy the same algebra. The two sets are, therefore, physically 
indistinguishable from each other. In retrospect we realize that relations (43) and (44) do not 
depend on the sign of $\kappa$ in Eqs. (37). This independence is important, since it restores 
the twofold degeneracy of the eigenvalues of the JT Hamiltonian.

The equivalence of the sets $\{A_j,\:A_j^{\dagger},\:N_j\}$ and $\{{\cal A},\:{\cal A}^{\dagger},
\:{\cal N}\}$ will now be exploited to construct the desired representation of the Hamiltonian on 
the space ${\cal V}_j$. Consider first the vibrational term ${\cal H}_v$, whose representation on 
${\cal U}_j$ is given by Eq. (36). To represent ${\cal H}_v$ on ${\cal V}_j$, we start by using 
the correspondence ${\cal N}\sim N_j$ and $d_{i\kappa}\sim c_{i\kappa}$. Thereby the matrix 
element $\langle i\kappa|{\cal N}|i\kappa\rangle$ is mapped on $(i\kappa|N_j|i\kappa)$, where 
$|i\kappa)=c_{i\kappa}^{\dagger}|0)_e$ and $|0)_e$ denotes the {\em electronic} vacuum in 
${\cal V}_j$. The representation of ${\cal H}_v$ on ${\cal V}_j$ is then obtained as

\begin{equation}
H_v^{(j)}/(\hbar\Omega)=\sum_i[(i\kappa|N_j|i\kappa)-1]=\sum_i(b_i^{\,\dagger}b_i+j-1/2) ,
\end{equation}
where Eq. (42b) has been used. To find the representation of the JT term, we must go back to Eq. 
(31) and exploit the correspondence ${\cal A}\sim A_j$. In this way the representation of 
${\cal H}_{JT}$ on ${\cal V}_j$ is found to be given by the expression

\begin{equation}
H_{JT}^{(j)}/(\hbar\Omega)= -g(A_j + A_j^{\dagger})
=-g\sum_{i\kappa}c_{i\kappa}^{\dagger}(A_i^{(j)}b_i+\mbox{H.c.})c_{i\kappa} ,
\end{equation}
where use has been made of Eq. (40a). For later purposes it proves more convenient to rewrite 
$H_{JT}^{(j)}$ in the form

\begin{equation}
H_{JT}^{(j)} = H_{JT}^{(0)}-g\hbar\Omega\sum_{i\kappa}c_{i\kappa}^{\dagger}
(P_+ B_i^{(j)}b_i+\mbox{H.c.})c_{i\kappa} ,
\end{equation}
where

\begin{mathletters}
\begin{equation}
H_{JT}^{(0)}=-g\hbar\Omega\sum_{i\kappa}c_{i\kappa}^{\dagger}(b_i^{\,\dagger}+b_i)c_{i\kappa}
\end{equation}
and

\begin{equation}
B_i^{(j)}=\left(1+\frac{2j}{b_i^{\,\dagger}b_i+1}\right)^{1/2}-1 . 
\end{equation}
\end{mathletters}\noindent
We mention in passing that expressions quite similar to those in Eqs. (45), (47) and (48) have 
been derived for the isolated $E\bigotimes e$ JT center by means of a rather different 
approach.$^{23}$ In the {\em unphysical} limit $j=0$ the JT interaction reduces to the 
displaced-oscillator term $H_{JT}^{(0)}$, since $B_i^{(j)}=0$ in this case.$^{22}$ The hopping 
term deserves some comment. 

In Sec. II it was pointed out that ${\cal H}_t$ does {\em not}, in general, commute with $\cal J$ 
implying that ${\cal H}_t$ may induce transitions between subspaces belonging to different 
eigenvalues $j$ of $\cal J$. However, from the vibrational term (45) it follows that the ground 
state belongs to $j=1/2$ and is separated from the state with $j=3/2$ by the energy $N\hbar\Omega$, 
which tends to infinity in the thermodynamic limit. This gives rise to a kind of selection rule, 
allowing transitions only within the ground-state manifold, and we may, therefore, regard 
${\cal H}_t$ as being restricted to the subspace ${\cal U}_{j=1/2}$. Hence, its representation on 
${\cal V}_{j=1/2}$ takes the form

\begin{equation}
H_t = -t\sum_{ia}\vec{c}_i{}^{\dagger}\cdot{\bf t}_a\cdot\vec{c}_{i+a} ,
\end{equation}
where the matrices ${\bf t}_a$ differ from the $\mbox{\boldmath $\tau$}_a$ in Eqs. (18b) only in 
the replacement of $\cal G$ by $G$.

Putting together our findings, the representation of the total Hamiltonian can now be easily written 
down. For the ground state ($j=1/2$), the result may be recast into the form

\begin{mathletters}
\begin{equation}
H = H_{QH} - g\hbar\Omega\sum_{i\kappa}c_{i\kappa}^{\dagger}
(P_+ B_i^{(j)}b_i+\mbox{H.c.})c_{i\kappa} ,
\end{equation}
where $B_i^{(j)}$ ($j=1/2$) is defined by Eq. (48b) and

\begin{equation}
H_{QH} = -t\sum_{ia}\vec{c}_i{}^{\dagger}\cdot{\bf t}_a\cdot\vec{c}_{i+a}
+ \hbar\Omega\sum_i b_i^{\,\dagger}b_i
- g\hbar\Omega\sum_{i\kappa}c_{i\kappa}^{\dagger}(b_i^{\,\dagger}+b_i)c_{i\kappa}
\end{equation}
\end{mathletters}\noindent
will be referred to as quasi-Holstein model. Although $H_{QH}$ and the standard Holstein 
model$^{24}$ have the same formal appearance, they differ from each other in two respects. First 
of all they differ in the hopping term, which is isotropic in the standard model and anisotropic in 
our case. A more subtle difference lies in the properties of the fermion operators: while those of 
the standard model commute with the boson operators, this is not the case with the $c_{i\kappa},\:
c_{i\kappa}^{\dagger}$ in Eqs. (50). The peculiar behavior of our fermion operators will, however, 
not entail any problems in the further analysis.

\subsection*{VI.~ VARIATIONAL TREATMENT}

The structure of the vibrational term (45) led us to conclude that the ground state of the 
$E\bigotimes e$ JT polaron belongs to $j=1/2$ and thus agrees with the well-known fact that the 
lowest state of the isolated $E\bigotimes e$ JT center belongs to the same quantum number.$^1$ 
Subsequently our Hamiltonian $H$ of Eqs. (50) will be subjected to a variational treatment, yielding
the (approximate) ground-state energy and the corresponding eigenvector for both the JT and the
Holstein polaron. From Eqs. (50) we see that the quasi-Holstein model results for $j=0$, while the 
JT case is obtained for $j=1/2$, allowing both types of polarons to be treated in a unified way.
A substantial simplification of the Hamiltonian may be achieved by an expansion of the square root 
in Eq. (48b) which, for $j=1/2$, converges for all eigenvalues of $b_i^{\,\dagger}b_i$. Keeping only 
terms linear in $j$, the expansion of $B_i^{(j)}$ becomes

\begin{equation}
B_i^{(j)}=j(b_i^{\,\dagger}b_i+1)^{-1}+O(j^2) .
\end{equation}
Since $B_i^{(j)}$ vanishes for $j=0$, we expect Eq. (51) to be reasonable for $j=1/2$. Indeed, for 
the isolated $E\bigotimes e$ JT center this turns out to be an excellent approximation,$^{23}$ and we 
shall use it in the subsequent analysis. 

One of the main ingredients of our treatment is a unitary operator (Jost transformation) having
the property that the transformed Hamiltonian assumes diagonal form with respect to the fermion 
momenta. This transformation, which has already proved its utility in various other polaron 
problems,$^{25}$ has the form

\begin{equation}
U=\exp\left(-i\sum_{i\kappa}c_{i\kappa}^{\dagger}{\bf Q\cdot R}_ic_{i\kappa}\right)
 =\sum_{i\kappa}c_{i\kappa}^{\dagger}U_ic_{i\kappa} ,
\end{equation}
where the last equality only holds on the single-particle Hilbert space. The operator

\begin{mathletters}
\begin{equation}
U_i=\exp(-i{\bf Q\cdot R}_i) ,
\end{equation}
where

\begin{equation}
{\bf Q}=\sum_{\bf q}{\bf q}\,b_{\bf q}^{\,\dagger}b_{\bf q}
\end{equation}
denotes the crystal momentum of the boson field, commutes with all $c_{i\kappa}$ and is a
translation operator for the Bose particles:

\begin{equation}
U_i^{\,\dagger}b_lU_i = b_{l-i} .
\end{equation}
\end{mathletters}

To better understand why the Jost transformation leads to a partial diagonalization of the
Hamiltonian in momentum space, we recall from Sec. II that the total crystal momentum
$\bf P=K+Q$, where ${\bf K}=\sum_{\bf k}{\bf k}\,\vec{c}_{\bf k}{}^{\dagger}\cdot\vec{c}_
{\bf k}$ denotes the fermion momentum and $\bf Q$ is given by Eq. (53b), is a conserved 
quantity. Using the alternative form $U=\sum_{\bf k}\vec{c}_{\bf k}{}^{\dagger}\cdot\vec{c}
_{\bf k+Q}$, one readily establishes the relation $U^{\,\dagger}{\bf P}U={\bf K}$, showing 
that in the transformed system $\bf K$ plays the role of the total crystal momentum. This 
necessarily implies that the transformed Hamiltonian becomes diagonal in the fermion momenta 
$\bf k$, the eigenvalues of $\bf K$, whereas the indices $\kappa$ remain unaffected. 

To obtain the transformed Hamiltonian $U^{\,\dagger}HU$, we use the last equality in Eq. (52) 
and property (53c). The calculation is straightforward and leads to the result

\begin{eqnarray}
\lefteqn{U^{\,\dagger}HU=-t\sum_{ia}U_a\vec{c}_i{}^{\dagger}\cdot{\bf t}_a\cdot\vec{c}_{i+a}
+\hbar\Omega\sum_ib_i^{\,\dagger}b_i}\hspace{0.3cm} \nonumber\\
 & & -g\hbar\Omega\sum_{i\kappa}c_{i\kappa}^{\dagger}[(1+P_+B_0^{(j)})b_0+\mbox{H.c.}]
c_{i\kappa} ,
\end{eqnarray}
where $b_0\equiv b_{i=0}$ and $B_0^{(j)}$ is given by Eq. (51). After Fourier transformation of 
the fermion operators the Hamiltonian assumes the expected form

\begin{mathletters}
\begin{equation}
U^{\,\dagger}HU=\sum_{\bf k}\sum_{\kappa\kappa'}c_{\bf k\kappa}^{\dagger}
H_{\bf k}^{\kappa\kappa'}c_{\bf k\kappa'} ,
\end{equation}
where

\begin{eqnarray}
\lefteqn{H_{\bf k}^{\kappa\kappa'}=-T_{\bf k-Q}^{\kappa\kappa'}+\delta_{\kappa\kappa'}
\hbar\Omega\{\sum_i b_i^{\,\dagger}b_i}\hspace{0.4cm} \nonumber\\
 & & -g[(1+P_+B_0^{(j)})b_0 + \mbox{H.c.}]\} 
\end{eqnarray}
and
\begin{equation}
T_{\bf k-Q}^{\kappa\kappa'}=t\sum_a t_a^{\kappa\kappa'}\exp[i{\bf (k-Q)\cdot R}_a] .
\end{equation}
\end{mathletters}\noindent

Our variational treatment is based on the displacement transformation

\begin{mathletters}
\begin{equation}
V=\sum_{\bf k\kappa}c_{\bf k\kappa}^{\dagger}V_{\bf k}c_{\bf k\kappa} ,
\end{equation}
defined on the single-particle Hilbert space, where the operator$^{26}$
\begin{equation}
V_{\bf k}=\exp\left[\frac{v_{\bf k}}{\sqrt{N}}\sum_{\bf q}(b_{\bf q}^{\,\dagger}-b_{\bf q})
\right] 
\end{equation}
\end{mathletters}\noindent
displaces the $b_{\bf q}$ according to the familiar formula $V_{\bf k}^{\dagger}b_{\bf q}
V_{\bf k}=b_{\bf q}+v_{\bf k}/\sqrt{N}$ ($N$ denotes the total number of sites). The unknown 
parameters $v_{\bf k}$ will be determined from the variational principle. The Hamiltonian 
then acquires the form

\begin{equation}
\tilde{H}\equiv (UV)^{\dagger}H(UV) = \sum_{\bf k}\sum_{\kappa\kappa'}c_{\bf k\kappa}^
{\dagger}V_{\bf k}^{\dagger}H_{\bf k}^{\kappa\kappa'}V_{\bf k}c_{\bf k\kappa'} ,
\end{equation} 
where $H_{\bf k}^{\kappa\kappa'}$ is given by Eq. (55b). Our further strategy may now be 
outlined as follows: as the first step, we introduce new fermion operators $f_{\bf k\kappa}$,
$f_{\bf k\kappa}^{\dagger}$ by means of the relation

\begin{equation}
f_{\bf k\kappa}^{\dagger}|0)=(\vec{c}_{\bf k}{}^{\dagger}\cdot {\bf L}_{\bf k})_{\kappa}|0)
=\sum_{\kappa'}L_{\bf k}^{\kappa'\kappa}c_{\bf k\kappa'}^{\dagger}|0) ,
\end{equation}
where ${\bf L}_{\bf k}$ is a unitary $2\times 2$ matrix, which will be specified below. In
terms of the new fermionic basis the Hamiltonian (57) takes the form

\begin{equation}
\tilde{H}=\sum_{\bf k}\sum_{\kappa\kappa'}f_{\bf k\kappa}^{\dagger}({\bf L}_{\bf k}^{\dagger}
V_{\bf k}^{\dagger}{\bf H}_{\bf k}V_{\bf k}{\bf L}_{\bf k})_{\kappa\kappa'}f_{\bf k\kappa'}
\end{equation}
and possesses the matrix elements

\begin{mathletters}
\begin{equation}
(0|f_{\bf k\kappa}\tilde{H}f_{\bf k\kappa'}^{\dagger}|0)=({\bf L}_{\bf k}^{\dagger}{\bf M}_
{\bf k}{\bf L}_{\bf k})_{\kappa\kappa'}=\sum_{\lambda\lambda'}(L_{\bf k}^{\lambda\kappa})^*
M_{\bf k}^{\lambda\lambda'}L_{\bf k}^{\lambda'\kappa'} ,
\end{equation}
where the expectation values

\begin{equation}
M_{\bf k}^{\lambda\lambda'}=(0|V_{\bf k}^{\dagger}H_{\bf k}^{\lambda\lambda'}V_{\bf k}|0)
\quad (\lambda,\lambda'=\pm 1)
\end{equation}
\end{mathletters}\noindent
will be functionals of $v_{\bf k}$. The matrix $\bf L_k$ is now fixed by the requirement

\begin{equation}
({\bf L}_{\bf k}^{\dagger}{\bf M}_{\bf k}{\bf L}_{\bf k})_{\kappa\kappa'}
={\cal E}_{\kappa}({\bf k})\delta_{\kappa\kappa'} ,
\end{equation}
which gives rise to two energy bands ${\cal E}_{\kappa}({\bf k})\:(\kappa=\pm 1)$. The 
variational parameters $v_{\bf k}$ are then obtained from the condition $\delta{\cal E}_-
({\bf k})/\delta v_{\bf k}=0$, where ${\cal E}_-({\bf k})$ denotes the band with the 
{\em lowest} energy. The approximate, normalized ground-state eigenvector has the form 

\begin{mathletters}
\begin{equation}
|\Phi_{\bf k-}) = UVf_{\bf k-}^{\dagger}|0) = Z_{\bf k}
\sum_{i\kappa}\sum_{n_i=0}^{\infty}C_{i\kappa,\bf k}(n_i)c_{i\kappa}^{\dagger}
\frac{(b_i^{\,\dagger})^{n_i}}{\sqrt{n_i!}}|0) ,
\end{equation}
where $Z_{\bf k}=N^{-1/2}\exp (-v_{\bf k}^2/2)$ and the coefficients are given by

\begin{equation}
C_{i\kappa,\bf k}(n_i)=\frac{(v_{\bf k})^{n_i}}{\sqrt{n_i!}}L_{\bf k}^{\kappa -}
\exp (i{\bf k\cdot R}_i) .
\end{equation}
\end{mathletters}

Thus, our first task is to evaluate the matrix elements (60b). A somewhat lengthy, but 
straightforward calculation gives the following results

\begin{mathletters}
\begin{eqnarray}
M_{\bf k}^{++} &=& -t_{\bf k}\sum_{a=x}^z\cos k_a + \hbar\Omega v_{\bf k}
(v_{\bf k}-2g)-2jg\hbar\Omega v_{\bf k}^{-1}\exp(-v_{\bf k}^2)\sinh v_{\bf k}^2 ,\\
M_{\bf k}^{+-} &=& -t_{\bf k}({\rm e}^{2i\pi/3}\cos k_x + 
{\rm e}^{-2i\pi/3}\cos k_y+\cos k_z) ,\\
M_{\bf k}^{--} &=& M_{\bf k}^{++} ,\quad M_{\bf k}^{-+} = (M_{\bf k}^{+-})^* ,
\end{eqnarray}
\end{mathletters}\noindent
where $j=0$ for the quasi-Holstein model, $j=1/2$ for the $E\bigotimes e$ JT polaron, and

\begin{equation}
t_{\bf k} = t\exp(-v_{\bf k}^2) .
\end{equation}
Having obtained the matrix $\bf M_k$, we readily find its eigenvalues ${\cal E}_{\kappa}
({\bf k})$,

\begin{equation}
{\cal E}_{\kappa}({\bf k})=-t_{\bf k}E_{\kappa}({\bf k})+\hbar\Omega v_{\bf k}(v_{\bf k}-2g)
-2jg\hbar\Omega v_{\bf k}^{-1}\exp(-v_{\bf k}^2)\sinh v_{\bf k}^2 ,
\end{equation}
where the quantities

\begin{equation}
E_{\kappa}({\bf k}) = \epsilon_0({\bf k})-\kappa\sqrt{\epsilon_1^2({\bf k})+\epsilon_2^2
({\bf k})} \quad (\kappa=\pm 1)
\end{equation}
denote the two $e_g$ bands in the absence of the JT coupling and
\begin{mathletters}
\begin{eqnarray}
\epsilon_0({\bf k}) &=& \cos k_x + \cos k_y + \cos k_z ,\\
\epsilon_1({\bf k}) &=& \frac{1}{2}(\cos k_x + \cos k_y - 2\cos k_z) ,\\
\epsilon_2({\bf k}) &=& \frac{\sqrt{3}}{2}(\cos k_x - \cos k_y) .
\end{eqnarray}
\end{mathletters}\noindent
The condition $\delta{\cal E}_-({\bf k})/\delta v_{\bf k}=0$ then yields a transcendental
equation for $v_{\bf k}$, which may be written as

\begin{mathletters}
\begin{equation}
v_{\bf k}=g\frac{1+2j\exp(-v_{\bf k}^2)F_{\bf k}}{1+(t_{\bf k}/\hbar\Omega)E_{\bf k}} ,
\end{equation}
where $E_{\bf k}\equiv E_-({\bf k})$ and

\begin{equation}
F_{\bf k} = \exp(-v_{\bf k}^2)-(2v_{\bf k}^2)^{-1}\sinh v_{\bf k}^2 .
\end{equation}
\end{mathletters}\noindent
To assess the range of validity of our variational approach, we shall now first investigate 
the limiting cases of weak and strong coupling.

\noindent
(i) {\em weak coupling}\,: $g\ll 1$

In this coupling range, Eqs. (68) possess the solution

\begin{mathletters}
\begin{equation}
v_{\bf k}=\frac{(1+j)g}{1+\gamma E_{\bf k}}+O(g^2) ,
\end{equation}
where the adiabaticity parameter

\begin{equation}
\gamma=t/\hbar\Omega .
\end{equation}
\end{mathletters}\noindent
If this is inserted into the expression for ${\cal E}_-({\bf k})$, Eq. (65), the result is

\begin{equation}
{\cal E}_-({\bf k})/\hbar\Omega = -\gamma E_{\bf k}-g^2\frac{1+2j}{1+\gamma E_{\bf k}} ,
\end{equation}
where terms of $O(j^2)$ have been excluded because of the expansion (51). To compare our
formula with exact analytical results for the Holstein polaron, we set $j=0$ and restrict 
ourselves to one dimension. At the $\Gamma$ point, Eq. (70) then reduces to

\begin{mathletters}
\begin{equation}
{\cal E}_-^{1D}(0)/\hbar\Omega = -\gamma - \frac{g^2}{1+\gamma} .
\end{equation}
This may now be compared with the result of weak-coupling perturbation theory,$^{27}$ which 
is valid for all $\gamma$:

\begin{equation}
{\cal E}_-^{1D}(0)/\hbar\Omega = -\gamma - \frac{g^2}{\sqrt{1+2\gamma}} .
\end{equation}
\end{mathletters}\noindent
While these two expressions agree for $\gamma\ll 1$, they start to diverge for larger
$\gamma$, and our formula (71a) gradually ceases to be a reasonable upper bound.$^{26}$ For 
the pure JT case ($\gamma=0,\,j=1/2$), Eq. (70) reduces to

\begin{equation}
{\cal E}_-({\bf k})/\hbar\Omega = -2g^2 ,
\end{equation}
which agrees with the result of perturbation theory.$^{21}$  

\noindent
(ii) {\em strong coupling}\,: $g\gg 1$

In this case, Eqs. (68) are solved by the expression

\begin{equation}
v_{\bf k} = g - \frac{j}{2g} + O(g^{-2}) ,
\end{equation}
which, after substitution into Eq. (65), leads to

\begin{mathletters}
\begin{equation}
{\cal E}_-({\bf k})/\hbar\Omega = - g^2 - j .
\end{equation}
For the Holstein polaron, strong-coupling perturbation theory$^{27}$ yields the result

\begin{equation}
{\cal E}_-({\bf k})/\hbar\Omega = - g^2 - \left(\frac{\gamma}{2g}\right)^2 ,
\end{equation}
\end{mathletters}\noindent
which agrees with our formula (74a) (for $j=0$) in the nonadiabatic limit $\gamma\ll 1$.
In the pure JT case ($j=1/2$), Eq. (74a) agrees with the strong-coupling expression in 
Ref. 21.

This concludes our discussion of the weak and strong coupling limits. Summarizing we may 
state that our variational treatment seems to work reasonably well in the nonadiabatic 
regime $\gamma\ll 1$, but becomes less reliable for larger $\gamma$.$^{26}$ The nonadiabatic 
regime might be relevant to the manganites. For, in the doping region considered in this 
work, the $t_{2g}$ core spins form an antiferromagnetic (G-type) spin background, leading to 
a strong suppression of hopping because of the double-exchange mechanism.$^9$   

Important characteristics of a polaron are its dispersion relation and effective mass. 
Subsequently these properties will be examined for both the quasi-Holstein model ($j = 0$) 
and the $E\bigotimes e$ JT case ($j = 1/2$) by means of a numerical evaluation of Eqs. (68). 
The result of such a calculation for $g = 1.5$ and $\gamma = 0.5$ is shown in Fig. 1, where 
the polaron dispersion relations ${\cal E}_-({\bf k})$ for $j = 0$ and $j = 1/2$ are depicted 
along a closed path of the cubic Brillouin zone (BZ). For comparison, the lower tight binding 
$e_g$ band ($g = 0$) is also shown in Fig. 1. Although the polaron bands are shifted to lower 
energies and have a smaller width in comparison with the $e_g$ band, as expected, the shapes 
of all three bands are very similar. In particular, all extrema of the dispersion curves are 
located at the same positions. Other prominent features of the bands are the extended flat 
minima between $\Gamma$ and $X$ and the absolute maxima at the $R$ point.

\hspace{6cm}\fbox{Figures 1 and 2}

The polaron effective mass $m^*$ is defined by the relation

\begin{displaymath}
\frac{1}{m^*} = \frac{1}{\hbar^2}\left(\frac{\partial^2{\cal E}_-({\bf k})}{\partial k^2}
\right)_{\bf k=0} ,
\end{displaymath}
where $k$ is the wave-vector component along some symmetry line of the BZ. Using Eq. (65)
we obtain the simple result

\begin{equation}
m_0/m^* = \exp (-v_{\bf k=0}^2) ,
\end{equation}
where

\begin{displaymath}
\frac{1}{m_0} = -\frac{t}{\hbar^2}\left(\frac{\partial^2E_{\bf k}}{\partial k^2}\right)_
{\bf k=0}
\end{displaymath}
denotes the inverse effective mass of the lower tight binding $e_g$ band. We mention that in
our derivation of Eq. (75) the $\bf k$ dependence of $v_{\bf k}$ has been properly taken into
account, the simplicity of the result being due to a cancellation of all terms involving 
$\partial^2v_{\bf k}/\partial k^2$. 

In Fig. 2 the mass ratio $m_0/m^*$ is plotted as a function of the coupling strength $g$. 
Although both polaron masses behave similarly, there is an unexpected crossover where, for
increasing $g$, the Holstein polaron starts to acquire a somewhat larger effective mass than 
the JT polaron. The effect is, however, not as dramatic as claimed by Takada.$^{29}$ For, in the 
strong-coupling limit, the polaron effective mass ratio $m_{JTP}^*/m_{HP}^*$ tends to the finite 
limit $\exp (-j)\approx 0.607$ for $j=1/2$, as follows from Eqs. (73) and (75). The results 
presented in Fig. 2 are in remarkably good agreement with recent quantum Monte Carlo data,$^{14}$ 
thus confirming our expectation that the proposed variational approach, restricted to the 
nonadiabatic regime, gives a fair account of polaronic properties over the whole coupling range.

\subsection*{VII.~ SUMMARY}

In this work a detailed account has been given of the analytic properties of the $E\bigotimes e$ 
JT polaron, consisting of a mobile $e_g$ electron linearly coupled to the local $e_g$ normal 
vibrations of a periodic array of octahedral complexes. The linear JT coupling implies the
existence of two operators, the angular momentum $\cal J$ and the parity $\cal K$, which commute 
with the JT part and are responsible for the twofold degeneracy of all JT eigenvalues. This 
degeneracy is lifted by the anisotropic hopping term, which does not commute with $\cal J$ and 
$\cal K$. The most interesting feature of our study is, however, the appearance of a close 
relationship between the JT problem and the Holstein model. Although such a connection has already 
been suspected to exist in the simpler $E\bigotimes b$ JT polaron,$^{11}$ it has never been 
explicitly demonstrated. This connection only emerges in a particular representation of the 
original problem, in which the Hamiltonian acquires an explicit dependence on the half-integral 
angular momentum quantum number $j$ and quite naturally decomposes into a Holstein term and a 
residual JT interaction. While the ground state of the JT polaron belongs to the sector $j=1/2$, 
the Holstein polaron is formally obtained for the unphysical value $j=0$. This is the optimal form 
of the Hamiltonian, which can be achieved by purely analytic means, allowing the JT and the 
Holstein polaron to be treated in a unified framework.

The Hamiltonian is then subjected to a variational treatment, yielding approximate ground-state
energies and eigenvectors for both types of polarons. Although the ground-state eigenvector is
explicitly given by Eqs. (62), its application to the calculation of physical properties is 
relegated to future work. Here we have restricted ourselves to the polaron dispersion relations and 
effective masses. As expected, the polaron bands are shifted to lower energies and have a smaller 
width in comparison with the bare $e_g$ band (see Fig. 1), but the shapes of all three bands are 
very similar. The dependence of the effective masses on the coupling strength $g$ is also similar for 
both polarons (see Fig. 2). There is, however, an unexpected crossover where the Holstein polaron 
starts to acquire a somewhat bigger mass than the JT polaron with increasing $g$. These results are 
in remarkably good agreement with recent quantum Monte Carlo data.$^{14}$ This seems to indicate that
our variational approach, restricted to the nonadiabatic regime for formal reasons, is fairly accurate.  
The nonadiabatic regime might be relevant to the manganites, for in the doping region considered in 
this work the core spins form an antiferromagnetic (G-type) spin background, leading to a strong 
suppression of hopping because of the double-exchange mechanism.$^9$

A more realistic model would have to include, at least, the intersite coupling of the normal
vibrations. This coupling is expected to contribute to the splitting of the degenerate JT ground state 
and to turn the Einstein phonons of the present work into optical phonons. Whether these coupling 
terms will give rise to additional and unexpected effects, remains to be seen and will be investigated
in future work.

\newpage
\appendix
\section{Direct proof of equivalence}

In Sec. V the equivalence of the operators $\cal A$ and $A_j$ has been proven in terms of their
algebraic properties. Here we intend to put forward a more direct proof by showing that the 
operators possess identical matrix elements, provided the states involved are related to each
other by the mapping prescription (39). We start by defining the basis vectors

\begin{mathletters}
\begin{eqnarray}
|i\kappa;m+2n_i,m\rangle &\equiv& d_{i\kappa}^{\,\dagger}|m+2n_i,m\rangle
\prod_{l\neq i}|0_l,0_l\rangle ,\\
|i\kappa;m+2n_i+1,m+1\rangle &\equiv& d_{i\kappa}^{\,\dagger}|m+2n_i+1,m+1\rangle
\prod_{l\neq i}|0_l,0_l\rangle ,
\end{eqnarray}
\end{mathletters}\noindent
where the notation is the same as in Eqs. (28) and $\kappa$ is fixed by Eq. (29c). The vectors
(A1a) and (A1b) are then eigenstates of the angular momentum operator $\cal J$ to the same 
eigenvalue $j=m+1/2$ and belong to the subspaces ${\cal U}_j^+$ and ${\cal U}_j^-$, respectively,
according to our definitions introduced at the end of Sec. IV. Given these basis states, the 
nonvanishing matrix elements of the operator $\cal A$, Eq. (30a), are readily evaluated by
means of the easily proven relations

\begin{mathletters}
\begin{eqnarray}
a_{ix}|m+2n_i,m\rangle &=& \sqrt{m+n_i}\,|m+2n_i-1,m-1\rangle ,\\
a_{iz}|m+2n_i,m\rangle &=& \sqrt{n_i}\,|m+2n_i-1,m+1\rangle ,
\end{eqnarray}
\begin{equation}
\Pi_{\kappa}|m+2n_i,m\rangle = 0 ,\quad \Pi_{-\kappa}|m+2n_i,m\rangle = |m+2n_i,m\rangle .
\end{equation}
\end{mathletters}\noindent
With the help of Eqs. (A2) we then find the expressions

\begin{eqnarray*}
{\cal A}|i\kappa;m+2n_i,m\rangle &=& \sqrt{2n_i}\,|i\kappa;m+2n_i-1,m+1\rangle ,\\
{\cal A}|i\kappa;m+2n_i+1,m+1\rangle &=& \sqrt{2j+2n_i+1}\,|i\kappa;m+2n_i,m\rangle ,
\end{eqnarray*}
and, hence, the only nonvanishing matrix elements of the operator $\cal A$ read:

\begin{mathletters}
\begin{eqnarray}
\langle i\kappa;m+2n_i-1,m+1|{\cal A}|i\kappa;m+2n_i,m\rangle &=& \sqrt{2n_i} ,\\
\langle i\kappa;m+2n_i,m|{\cal A}|i\kappa;m+2n_i+1,m+1\rangle &=& \sqrt{2j+2n_i+1} .
\end{eqnarray}
\end{mathletters}

Our assertion is that the matrix elements of the operator $A_j$, defined by Eqs. (40), are the 
same as those on the right side of Eqs. (A3), provided the states are chosen as prescribed by 
Eq. (39). To prove our claim, we first introduce the new basis vectors  

\begin{mathletters}
\begin{eqnarray}
|i\kappa;2n_i) &\equiv& c_{i\kappa}^{\dagger}\frac{(b_i^{\,\dagger})^{2n_i}}
{\sqrt{(2n_i)!}}|0) ,\\
|i\kappa;2n_i+1) &\equiv& c_{i\kappa}^{\dagger}\frac{(b_i^{\,\dagger})^{2n_i+1}}
{\sqrt{(2n_i+1)!}}|0) ,
\end{eqnarray}
\end{mathletters}\noindent
where we have adopted the notation of Eqs. (37). Here, in contrast to the vectors (A1), the
quantum number $\kappa$ needs no longer to be fixed, but may be arbitrarily set equal to $1$ or 
$-1$. Since, by definition, the vectors (A4a) and (A4b) are elements of the subspaces 
${\cal V}_j^+$ and ${\cal V}_j^-$, respectively, the mapping prescription (39) requires the 
following one-to-one correspondence to exist between the vectors (A1) and (A4):

\begin{mathletters}
\begin{eqnarray}
|i\kappa;m+2n_i,m\rangle &\leftrightarrow& |i\kappa;2n_i) ,\\
|i\kappa;m+2n_i+1,m+1\rangle &\leftrightarrow& |i\kappa;2n_i+1) ,\\
|i\kappa;m+2n_i-1,m+1\rangle &\leftrightarrow& |i\kappa;2n_i-1) .
\end{eqnarray}
\end{mathletters}\noindent 
By means of these relations the matrix elements (A3) are then mapped onto the following
expressions
 
\begin{mathletters}
\begin{eqnarray}
(i\kappa;2n_i-1|A_j|i\kappa;2n_i) &=& \sqrt{2n_i} ,\\
(i\kappa;2n_i|A_j|i\kappa;2n_i+1) &=& \sqrt{2j+2n_i+1} ,
\end{eqnarray}
\end{mathletters}\noindent
whose validity is explicitly verified with the help of Eqs. (A4) and (40). This proves the
equivalence of the operators $\cal A$ and $A_j$, as far as the basis states are concerned. Owing 
to the completeness of these states, however, relations (A3) and (A6) suffice to extend the proof 
to arbitrary vectors of the spaces ${\cal U}_j$ and ${\cal V}_j$, provided these vectors are 
related to each other by Eq. (39). This completes our direct proof of the equivalence of the
operators $\cal A$ and $A_j$.

\newpage
{\bf Figure Captions}

\figure{Fig. 1. Polaron dispersion curves ${\cal E}_-({\bf k})/\hbar\Omega$, Eq. (65), for $j=0$ 
(Holstein polaron) and $j=1/2$ ($E\bigotimes e$ JT polaron). The lower tight binding $e_g$ band 
($g=0$) is also shown for comparison. Symmetry points of the BZ are designated as in Ref. 28.}

\figure{Fig. 2. Inverse effective masses of the Holstein polaron (dashed line) and the 
$E\bigotimes e$ JT polaron (solid line) as functions of the coupling strenght $g$.}


\begin{references}
\bibitem{20} For $N$ sites, the degeneracy of the energy levels $E_{nj\kappa}^{\,0}$ is given 
by the formula (see Ref. 19)

\begin{displaymath}
d_n = 2(n+1){n+N-1\choose N-1} ,
\end{displaymath}
where the factor $2(n+1)$ originates from the fact that $\kappa$ ranges over two and $j_{\kappa}$ 
over $n+1$ values, but $E_{nj\kappa}^0$ does not depend on these quantum numbers. The binomial 
coefficient gives the number of ways the relation $n=\sum_i n_i$ can be fulfilled in terms of the 
natural numbers $n_i$.
\end{references}
\end{document}